\documentclass[12pt]{article}
\usepackage{amsgen,amsthm,amsmath,amstext,amsbsy,amsopn,amssymb,latexsym, bbm}
\usepackage{mathtools}
\usepackage[usenames]{color}
\usepackage{array}
\usepackage{multirow}
\usepackage{bm}
\usepackage{amssymb}
\usepackage{mathabx}
\usepackage{natbib}
\usepackage{booktabs}
\usepackage{wrapfig}
\usepackage[ruled,vlined,linesnumbered,noresetcount]{algorithm2e}
\usepackage{hyperref}
\usepackage{pifont}
\usepackage{float}
\usepackage{tikz}
\usepackage{setspace}
\usepackage{graphicx} 
\usepackage[font=small,labelfont=bf]{caption} 
\usepackage[subrefformat=parens]{subcaption}
\usepackage{enumerate} 
\usepackage{pgfplots}
\pgfplotsset{width=10cm,compat=1.9}
\usepackage{amsthm}
\usepackage{authblk}

\newcommand{\citeg}[1]{\citep[e.g.,][]{#1}}
\usepackage{authblk}

\def\trans{^{\scriptscriptstyle \sf T}}

\newcommand{\HH}{{\mathcal{H}}}

\newcommand{\PP}{{\mathbf{P}}}

\newcommand{\Tbb}{\mathbb{T}}

\def\Pbb{\mathbb{P}}
\def\supP{^{\scriptscriptstyle (\Pbb)}}
\def\supQ{^{\scriptscriptstyle (\Qbb)}}

\def\supQbb{^{\scriptscriptstyle (\Qbb)}}

\def\bbetatilde{\widetilde{\bbeta}}

\def\betahat_1{\widehat{\beta}_1}

\def\Pbb{\mathbb{P}}
\def\Ebb{\mathbb{E}}

\usepackage{mathtools}

\def\supl{^{\scriptscriptstyle \sf \text{\tiny(}l\text{\tiny)}}}
\def\trans{^{\intercal}}

\def\bbeta{\beta}
\def\boeta{\boldsymbol{\boeta}}

\def\bbetahat{\widehat{\bbeta}}
\def\bbetatilde{\widetilde{\bbeta}}

\DeclareMathOperator*{\argmin}{arg\,min}

\definecolor{jcolor}{RGB}{041,122,000}
\definecolor{darkblue}{RGB}{000,000,150}
\definecolor{darkred}{RGB}{100,000,000}
\definecolor{purple}{RGB}{200,000,200}

\def\supinit{^{\scriptscriptstyle \sf (ini)}}
\def\Pbb{\mathbb{P}}

\def\supL{^{\scriptscriptstyle (L)}}

\def\Psc{\mathcal{P}}

\def\supl{^{\scriptscriptstyle (l)}}

\def\Qbb{\mathbb{Q}}

\def\Csc{\mathcal{C}}
\def\Dsc{\mathcal{D}}

\def\bbeta{\boldsymbol{\beta}}
\def\bbetahat{\boldsymbol{\widehat{\beta}}}
\def\BBhat{\boldsymbol{\widehat{B}}}

\def\subX{_{\scriptscriptstyle X}}

\def\Qbb{\mathbb{Q}}
\def\Pbb{\mathbb{P}}

\def\supTbb{^{\scriptscriptstyle (\Tbb)}}

\def\BBhat{\widehat{\BB}}

\def\betahat{\widehat{\beta}}

\def\Csc{\mathcal{C}}

\def\BB{\mathbf{B}}

\def\XX{\mathbf{X}}
\def\YY{\mathbf{Y}}

\def\Psc{\mathcal{P}}

\def\supl{^{\scriptscriptstyle (l)}}

\def\supL{^{\scriptscriptstyle (L)}}

\def\supmixT0{^{\mathrm{mixT0}}}

\def\subX{_{\scriptscriptstyle \sf X}}
\def\subYX{_{\scriptscriptstyle \sf Y|X}}

\let\OLDthebibliography\thebibliography
\renewcommand\thebibliography[1]{
  \OLDthebibliography{#1}
  \setlength{\parskip}{0pt}
  \setlength{\itemsep}{0pt plus 0.3ex}
}

\newcommand{\blind}{0}
\expandafter\def\expandafter\normalsize\expandafter{%
    \normalsize%
    \setlength\abovedisplayskip{5pt}%
    \setlength\belowdisplayskip{6pt}%
    \setlength\abovedisplayshortskip{-8pt}%
    \setlength\belowdisplayshortskip{2pt}%
}

\allowdisplaybreaks
\begin{document}
\def\spacingset#1{\renewcommand{\baselinestretch}%
{#1}\small\normalsize} \spacingset{1}

\doublespacing

\if1\blind
{
  \title{\bf Adversarial Drift-Aware Predictive Transfer: Toward Durable Clinical AI}
  \author{}
  \date{}
  \maketitle
} \fi

\if0\blind
{
\title{\bf Adversarial Drift-Aware Predictive Transfer: Toward Durable Clinical AI}
\date{}
{
\author[*]{
Xin Xiong$^1$, Zijian Guo$^2$, Haobo Zhu$^3$,  Chuan Hong$^3$, Jordan W Smoller$^{4,5}$, Tianxi Cai$^{1,6}$, Molei Liu$^{7,8}$}
\affil[1]{Department of Biostatistics, Harvard T.H. Chan School of Public Health, USA}
\affil[2]{Center for Data Science, Zhejiang University, China.}
\affil[3]{Department of Biostatistics \& Bioinformatics, Duke University, USA.}
\affil[4]{Department of Psychiatry, Harvard Medical School, USA.}
\affil[5]{Department of Psychiatry, Massachusetts General Hospital, USA.}
\affil[6]{Department of Biomedical Informatics, Harvard Medical School.}
\affil[7]{Department of Biostatistics, Peking University Health Science Center, China.}
\affil[8]{Beijing International Center for Mathematical Research, Peking University, China.}
\affil[*]{Correspondence to: moleiliu@bjmu.edu.cn}
}
\maketitle
} \fi

\begin{abstract}

Clinical AI systems frequently suffer performance decay post-deployment due to temporal data shifts, such as evolving populations, diagnostic coding updates (e.g., ICD-9 to ICD-10), and systemic shocks like the COVID-19 pandemic. Addressing this ``aging'' effect via frequent retraining is often impractical due to computational costs and privacy constraints. To overcome these hurdles, we introduce Adversarial Drift-Aware Predictive Transfer (ADAPT), a novel framework designed to confer durability against temporal drift with minimal retraining. ADAPT innovatively constructs an uncertainty set of plausible future models by combining historical source models and limited current data. By optimizing worst-case performance over this set, it balances current accuracy with robustness against degradation due to future drifts. Crucially, ADAPT requires only summary-level model estimators from historical periods, preserving data privacy and ensuring operational simplicity. Validated on longitudinal suicide risk prediction using electronic health records from Mass General Brigham (2005--2021) and Duke University Health Systems, ADAPT demonstrated superior stability across coding transitions and pandemic-induced shifts. By minimizing annual performance decay without labeling or retraining future data, ADAPT offers a scalable pathway for sustaining reliable AI in high-stakes healthcare environments.

\end{abstract}

\noindent%
{\it Keywords:}  Model Degradation; Concept Drift; Generalizability; Distributionally Robust Optimization; Electronic Health Record; Suicide Risk Prediction.

\doublespacing 

\section{Introduction}


The integration of artificial intelligence (AI) into healthcare has catalyzed advances in early diagnosis, risk stratification, and personalized interventions. AI-driven decision support tools (clinical AI systems) are increasingly approved by the U.S. Food and Drug Administration (FDA) as medical devices \citep{muehlematter2023fda}. Nevertheless, ensuring safety and effectiveness after deployment remains challenging \citep{abulibdeh2025illusion}. A critical limitation in real-world application is the temporal degradation of model performance. Extensive studies indicate that most clinical AI models experience substantial effectiveness reductions within years of deployment \citep{vela2022temporal,fernandez2025unsupervised}. This ``aging'' phenomenon is particularly pronounced in clinical settings due to the dynamic nature of healthcare ecosystems, where data distributions shift continuously in response to evolving clinical practices, updates in diagnostic criteria, demographic changes, and external disruptions such as pandemics.

Clinical AI models are particularly susceptible to performance degradation due to two primary, and often intertwined, forms of temporal drift \citep{2025Statistically}. First, covariate shift arises when the distribution of input features changes over time, driven by factors such as evolving patient demographics, modifications in EHR documentation practices, or transitions in coding systems (e.g., from ICD-9 to ICD-10) \citep{cartwright2013icd,fernandez2025unsupervised}. Second, concept drift occurs when the relationship between features and outcomes shifts, such as when the determinants of suicide risk evolve during societal disruptions like the COVID-19 pandemic \citep{sheu2023efficient}, or when updates to clinical assessment guidelines redefine outcome criteria. Because changes in clinical practice or coding standards can simultaneously influence both input distributions and outcome definitions, covariate and concept shifts often emerge in tandem, compounding the challenge of model reliability over time.

Despite the prevalence of these challenges, existing domain adaptation and transfer learning methods often fall short in ensuring long-term durability. Whether addressing covariate shift \citeg{gretton2009covariate,reddi2015doubly,liu2020doubly} or concept drift--which typically necessitates labeled data from the current target \citeg{li2020transfer,tian2023transfer}--these approaches primarily optimize for the present distribution, critically overlooking robustness against degradation caused by future drift. Consequently, maintaining model reliability often defaults to a naïve solution: continuously collecting forthcoming data to frequently retrain or update the system. However, this strategy is rarely feasible in clinical practice due to the prohibitive costs of repeated expert annotation, computational demands, and regulatory hurdles. Thus, there is a notable absence of systematic approaches designed to not only adapt to the current target but also proactively guard against post-deployment deterioration \citep{muralidharan2024scoping}.

In parallel, distributionally robust optimization (DRO) has been developed to enhance generalizability of models against unseen distributional shifts and uncertainties \citeg{rahimian2019distributionally,duchi2021learning}. It aims to optimize worst-case performance over a pre-specified uncertainty set of possible ``future'' distributions. As a special class of DRO, group DRO and maximin methods \citep{MeinshausenBuhlmann2015,sagawa2019distributionally} define uncertainty sets based on pre-defined subgroups, e.g., demographic groups or data in different institutions. These methods can naturally address heterogeneity across multiple data sources or time windows. However, their direct application often yield overly conservative models that sacrifice accuracy on the current target population, which limits their clinical utility. While some recent methods \citeg{xiong2023distributionally,zhan2024domain,mo2024minimax} aim to solve this dilemma, their applicability remains limited to linear models and has not been extensively investigated under temporally evolving data regimes, such as those driven by evolving clinical practices, ICD coding transitions or pandemic-induced concept shifts. In addition, implementations of many existing DRO methods require access to individual-level data from all sources for training, raising concerns on costs of implementation and privacy. 

\begin{figure}[htb!]
    \centering    \includegraphics[width=1\linewidth]{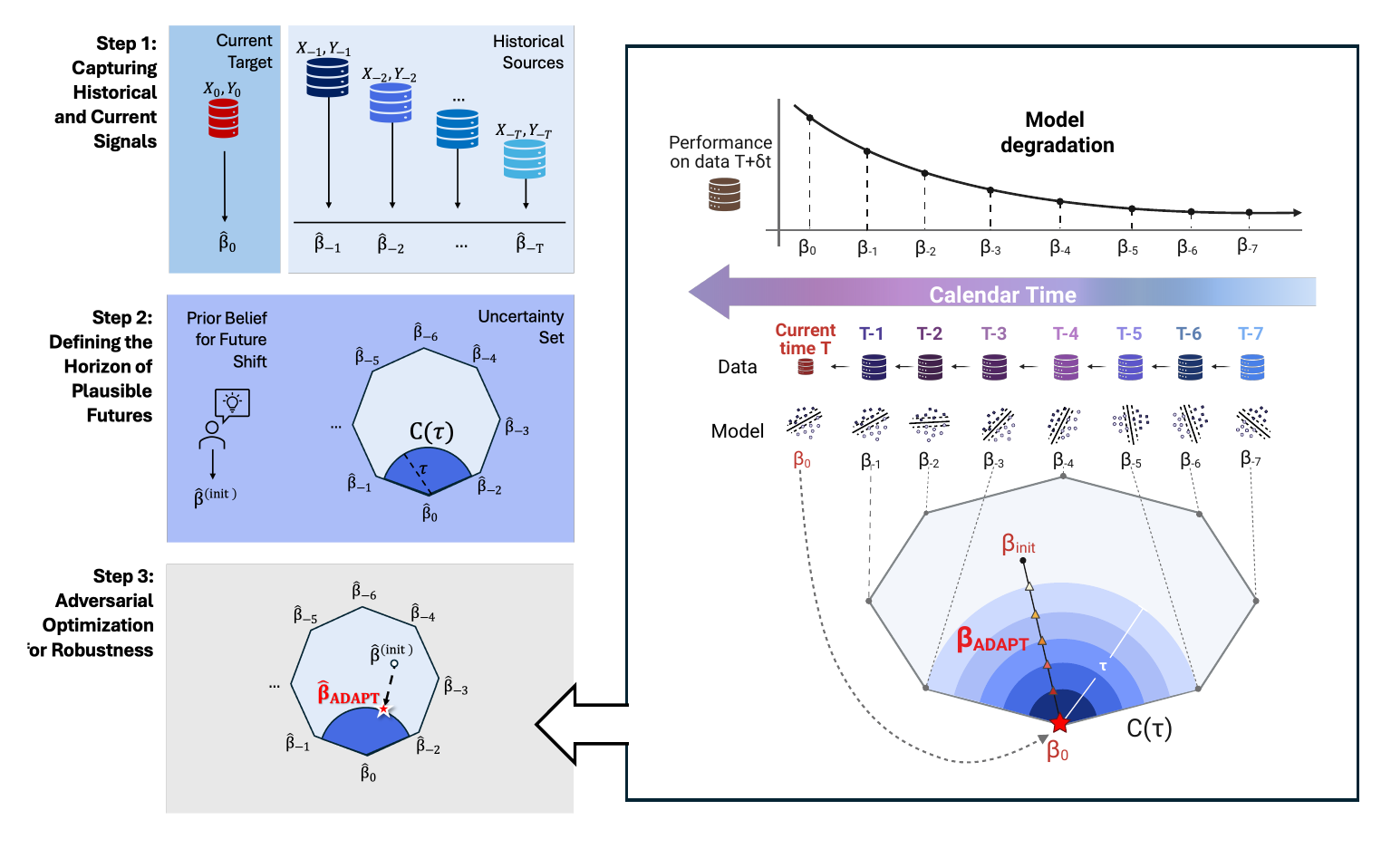}
    \caption{(Left panel) Three-step ADAPT algorithm: (1) fit period-specific models for the current target year and prior years to obtain a set of candidate models; (2) use the current and historical models to characterize the range of plausible futures by specifying an uncertainty set over candidate models that likely represents future drift; and (3) solve an adversarial robustness optimization problem over this region to produce the ADAPT model. (Right panel) Expanded view of Step~3 with the degradation motivation: model performance decays as calendar time moves away from the training period (older $\bbeta_t$ degrade on later data), and ADAPT selects $\bbeta_{\mathrm{ADAPT}}$ as a present-time robust solution anchored at $\bbeta_0$ that hedges against plausible near-future drift within the uncertainty set $C(\tau)$, improving stability against degradation.}
    \label{fig:flowchart}
\end{figure}

To address the challenges reviewed above, we develop Adversarial Drift-Aware Predictive Transfer (ADAPT), a novel framework designed to maintain the robustness of clinical risk models and extend their lifespan without the need for future retraining. Outlined in Figure \ref{fig:flowchart}, ADAPT moves beyond reactive updates by leveraging adversarial optimization to reconcile historical knowledge with emerging trends, ensuring predictions remain stable for future patients. Its core mechanism introduces two practical innovations. First, instead of overfitting to the present, ADAPT constructs an uncertainty set of plausible future scenarios derived as mixtures of historical and current patterns, and optimizes the model to withstand the worst-case drift within this set. This strategy enables a critical trade-off, maintaining high accuracy on the current target population while securing robustness against potential future degradation. Second, to ensure scalability and privacy, ADAPT utilizes a ``one-step'' analytical approximation. This allows the model to be updated efficiently using only summary-level statistics from historical sites, bypassing the computational burden and privacy risks associated with sharing raw patient data or training complex models from scratch.

ADAPT is validated on extensive synthetic data studies and a real-world study on suicide risk prediction using electronic health record (EHR) data from Mass General Brigham (MGB) Research Patient Data Registry (RPDR) mental health patients collected during 2005--2021 spanning coding system transitions and the pandemic, with the longitudinal EHR cohort from Duke University Health Systems (Duke) as an external validation. Compared to state-of-the-art methods, our approach demonstrates advantages in (i) achieving accurate prediction on current target data through the knowledge transfer from historical information; (ii) maintaining stable long-term performance and robustness against model degradation in future populations; (iii) identifying key risk predictors generalizable across different time periods. ADAPT establishes a scalable framework for maintaining AI reliability in dynamic clinical environments. By integrating adversarial robustness, privacy preservation, and minimal retraining, it addresses a critical translational gap in healthcare AI.

\section{Method}

\subsection{The ADAPT Framework}\label{sec:method:alg}

\paragraph{Overview.}
Ensuring the longevity of clinical AI requires models that are not only accurate for the current patient population but also robust against inevitable future shifts of the population. Standard training methods often overfit to the present, leading to rapid degradation. Our primary objective is to achieve robust prediction for future target populations that may differ from the current data but share underlying structural similarities. The observed current data provide the most recent and directly relevant representation of the population of interest, yet future populations are likely to continue evolving due to ongoing temporal drift. 

We develop the Adversarial Drift-Aware Predictive Transfer (ADAPT) framework, designed to produce models robust to temporal drift by exploiting both the heterogeneity among historical source populations and the characteristics of the most recent (current) target population. It operates on a fundamental premise: while the future is unknown, it likely exists within a continuum of mechanisms observed in the past and present. By mathematically constructing a set of plausible future scenarios and optimizing for the ``worst-case'' among them, ADAPT creates a risk model that not only stabilizes predictions for the current target but also anticipates and generalizes to plausible distributional shifts that may emerge in the future. The framework operates in three streamlined steps, requiring only summary-level statistics from historical data, thereby preserving patient privacy and minimizing computational overhead.

\paragraph{Step 1: Capturing Historical and Current Signals.}
The first stage involves extracting knowledge from available data sources. We assume access to multiple historical datasets (sources) collected over time, as well as a smaller, more recent dataset representing the current target population. Rather than pooling all data (which dilutes recent trends) or training solely on new data (which ignores historical stability), we first estimate separate baseline models for each time period. In specific, we utilize generalized linear models (GLMs) to derive coefficient vectors, denoted as $\{\bbeta^{(1)}, \ldots, \bbeta\supl\}$ and $\bbeta\supQbb$, from the historical sources $l=1,\ldots,L$ and the current target $\supQbb$, respectively. Note that although we consider the GLM here, our framework accommodates more sophisticated neural network methods for which the predictors can be naturally taken as the last one or few layers of some pretrained neural network.

Importantly, to prevent overfitting and ensure proper construction, we randomly partition the current target data into two subsets: an \textit{estimation set} used to train the initial current-target model, and a \textit{held-out evaluation set} reserved to validate the plausibility of mixture models as introduced in the subsequent step.

\paragraph{Step 2: Defining the Horizon of Plausible Futures.}
A core challenge in robust AI is defining what a ``plausible'' future shift looks like. We address this by constructing an \textit{uncertainty set}, a mathematical boundary containing potential future model configurations. We postulate that the future model parameters will likely reside within the {\em convex combination} (or {\em weighted mixture}) of the historical source models and the current target model. This conceptualizes the future as a complex recombination of past mechanisms (e.g., prior coding practices or demographic mixtures) and current trends. However, not all mixtures are likely. We refine this set by retaining only those model combinations that perform adequately well on the held-out evaluation set from Step 1. This process effectively filters out mixture models that are completely obsolete while retaining those that still hold predictive value. The size of this uncertainty set is governed by a carefully chosen tuning parameter that balances two competing needs: fitting the current data tightly (accuracy) versus allowing for deviation to account for drift (robustness). 


Note that our convex combination construction serves as a regularizing device rather than a strict assumption. While the mixture relationship needs not hold exactly, it encodes a prior belief that the target model $\bbeta\supQbb$ lies near the subspace spanned by $\{\bbeta^{(1)}, \ldots, \bbeta\supl\}$. This constraint can also help stabilize estimation under limited target data.

\paragraph{Step 3: Adversarial Optimization for Robustness.}
In the final step, ADAPT determines the optimal final model $\bbeta_{\text{ADAPT}}$. Instead of simply minimizing the error on the current data or any one specific plausible future distribution constructed in Step 2, we employ a minimax (minimize the maximum loss) optimization strategy. ADAPT searches for the model that minimize the prediction error under the \textit{worst-case} scenario found within our uncertainty set. Conceptually, the algorithm asks: ``Among all plausible future shifts defined in Step 2, which one would cause the most failure, and how do we adjust our model today to mitigate that failure?'' While adversarial training is typically computationally expensive or intractable, ADAPT utilizes a second-order approximation around the current target model. This technical innovation allows us to solve the optimization problem analytically in a single step, without the need for complex iterative training. The result is a closed-form estimator that is computationally efficient and statistically stable.

\paragraph{Summary of Advantages.}
By synthesizing historical knowledge with recent data through this adversarial lens, ADAPT offers the following three distinct clinical advantages. \textbf{Drift Robustness:} It anticipates future shifts rather than reacting to them, reducing the frequency of required retraining. \textbf{Accuracy on Current Target:} Compared to typical adversarial learning methods, ADAPT elaborates on the set of plausible future targets to avoid over-conservativeness and ensure accuracy on the current target. \textbf{Data Privacy:} It requires only the model coefficients from historical sites or time periods, not the transportation of patient-level data. \textbf{Operational Efficiency:} The ``one-step'' solution in Step 3 is essentially instantaneous compared to more complex learning retraining, making it scalable for real-time deployment in dynamic health systems. The mathematical details of our setup and algorithm are provided in Section \ref{sec: supp a} of the Supplementary Material.

\subsection{Validation Studies}

We conducted extensive validation and comparison studies on the proposed \textbf{ADAPT} algorithm using both synthetic experiments and a real-world case study. The synthetic data experiments were designed to systematically assess ADAPT’s robustness and generalizability under controlled conditions of temporal drift, while the real-world evaluation focused on developing a generalizable suicide risk prediction model using EHR data from Mass General Brigham (MGB) and evaluating its performance in MGB as well as an external cohort from Duke University Health Systems (Duke). Together, these studies allow us to benchmark ADAPT against existing domain adaptation and drift-robust methods, demonstrating its stability and predictive advantage in diverse scenarios.

\subsubsection{Validation Using Synthetic Data}

\paragraph{Data Generation.}

To rigorously evaluate ADAPT under realistic temporal dynamics, we generated binary-outcome datasets across discrete time periods $l \in \{1, \dots, L=15\}$ using a logistic model with $p=100$ standard normal predictors. The time-varying coefficient vectors $\bbeta\supl$ were designed to encapsulate both smooth autoregressive evolution (simulating gradual temporal heterogeneity) and abrupt distributional shifts. Initialized with sparse seed vectors, the coefficients evolved via a weighted mixing of prior models subject to stochastic shocks, simulating gradual clinical drift. To emulate a sudden systemic shift (e.g., a coding transition), we introduced a global perturbation at $l=8$, where the magnitude of this abrupt change is controlled by the perturbation level $p_{\mathrm{perturb}}$. Accommodating two types of distributional shifts, our construction can provide a rigorous testbed for evaluating ADAPT’s ability to handle degradation. Details of our data generation procedure are provied in Section \ref{sec:supp:sim} of Supplementary Material.

\paragraph{Simulation Configurations.} 
We varied two key factors to evaluate model performance under different 
conditions of data availability and unanticipated shocks: (i) the 
current-to-historical data ratio $\rho$, and (ii) the perturbation level 
$p_{\mathrm{perturb}}$. Here, $\rho$ denotes the ratio of the current 
training sample size to the per-period historical size ($N=2000$), with 
current size $n=\rho N$. To examine the effect of data availability, we 
varied $\rho$ from $0.1$ to $1$, gradually increasing access to current 
data during training. To 
assess resilience to abrupt distributional shifts, we varied 
$p_{\mathrm{perturb}}$ from $0$ (no perturbation) to $0.9$ (large 
perturbation), fixing $\rho=0.2$.  
This design allows us to disentangle the effects of limited current data 
and different degrees of perturbation, thereby assessing each method’s 
robustness to both gradual temporal drift and sudden global shifts in the 
data-generating process.

\paragraph{Benchmarks and Evaluation Metric.} 

We compare \textsc{ADAPT} against several benchmark methods: 
(i) a \textbf{Target-only} $\ell_1$-regularized logistic model trained 
exclusively on the current period $T_0$, 
(ii) a \textbf{Pooled} $\ell_1$-regularized logistic model trained on all 
data available up to and including $T_0$, 
(iii) \textbf{TransGLM} \citep{tian2023transfer}, which jointly estimates source and target models with penalties that shrink the target toward similar sources, and (iv) the distributionally robust \textbf{Maximin} estimator \citep{MeinshausenBuhlmann2015}, which combines source models estimated from each period to guard against worst-case distributional shifts. Since Maximin was originally designed for squared-loss regression, we adapt it to logistic models via a second-order local approximation of the negative log-likelihood relative to the null model around the current target. This yields an objective of the same quadratic form as \eqref{eq:adapt_est}, but with the simplex constraint (i.e., $\Csc(\infty)$) on the aggregation weights and with $\bbeta\supinit$ fixed to the zero vector. Performance is evaluated on the current period $T_0$ to assess 
transferability, and on all future periods with $l>T_0$ to assess 
generalizability under temporal shift. We quantify the performance based on the area under the receiver operating characteristic curve (AUC).

\subsubsection{Real World Case Study of Suicide Risk Prediction Model Degradation}

Suicide is a major public health concern, with an estimated 727,000 deaths worldwide in 2021 \citep{world2025suicide} and more than 45,000 annually in the United States, where the age-adjusted rate exceeds 13 per 100,000 \citep{Ahmad_2024}. It is the second leading cause of death among individuals aged 10–24, and rates have continued to rise over time \citep{Ormiston_2024,hua2024suicide}. In addition to these fatalities, approximately 1.5 million people attempt suicide annually in the United States, underscoring the urgent need for early identification and prevention efforts. Timely risk stratification is critical to prevent deaths \citep{kiran2024controlled}, yet suicide rates, risk factors, and the effectiveness of prevention efforts vary substantially across racial and ethnic groups, clinical settings, and demographic subpopulations \citep{coppersmith2024heterogeneity}. Point-of-care suicide risk prediction models leveraging EHRs offer a promising strategy to capture longitudinal trajectories and provide continuously updated risk estimates, but their reliability is undermined by temporal shifts (e.g., coding changes, evolving prevalence) and heterogeneity across health systems (e.g., data standards, clinical practices). Models trained on earlier data may degrade after deployment, limiting clinical utility \citep{walsh2025risk}. We examine whether ADAPT can learn stable and temporally robust suicide risk models by adaptively borrowing strength from heterogeneous historical EHR data.

\paragraph{Description of the Study Cohorts.}
The primary study cohort was derived from the Mass General Brigham (MGB) Research Patient Data Registry (RPDR), a centralized EHR repository encompassing more than 7 million patients across over eight affiliated hospitals, including Massachusetts General Hospital and Brigham and Women’s Hospital \citep{sheu2023ai}. 
We included patients with at least one mental health–related diagnosis (PheCodes 295–306) recorded between January~1,~2005 and December~31,~2022, requiring a minimum of six months of follow-up. 
PheCodes, which aggregate related ICD codes into clinically meaningful diagnostic groups \citep{denny2013systematic}, were used both to define the mental health cohort and to construct diagnosis-based features. 
For each eligible patient, a single ``landmark visit'' was randomly selected between the date of the first mental health diagnosis and the 75th percentile of the patient’s observation window. 
The prediction target was suicide risk within one year following the landmark visit, identified using ICD-9 and ICD-10-CM codes corresponding to incident suicide attempts \citep{suicide-icd910, barak2017predicting}.

To comprehensively evaluate model generalizability, we extended validation beyond a single health system by incorporating a second independent EHR cohort from the Duke University Health Systems (Duke). The Duke database contains longitudinal records for more than 6 million patients from 2015 through 2022. We constructed the Duke cohort using the same inclusion criteria and feature definitions as applied to the MGB RPDR cohort, ensuring consistency in both outcome ascertainment and feature representation. 
This dual-cohort design enables assessment of ADAPT’s performance across two complementary dimensions of external validity: 
(i)~\emph{temporal generalization}, by examining model robustness across calendar years within each institution; and 
(ii)~\emph{cross-institutional generalization}, by evaluating model transportability between distinct healthcare systems with differing patient populations and documentation practices. Details of cohort characteristics, including sample sizes by calendar year at both MGB and Duke, are provided in Section \ref{sec: supp demo} of the Supplementary Materials.

\paragraph{Feature Construction.}
To identify predictive features, we used the Online Narrative and Codified feature Extraction (ONCE) framework, which applies knowledge graph–based representation learning to derive clinically meaningful variables from EHR data \citep{komap}. ONCE identified 103 suicide-related features mapped to standardized clinical ontologies—diagnoses (ICD codes via PheCodes), medications (RxNorm), procedures (CCS), and laboratory tests (LOINC)—that translate raw EHR data into modeling-ready representations \citep{huang2025advancing}. Each feature was represented by the $\log(1+x)$-transformed count within the one-year baseline window preceding the landmark visit. Demographic variables (age at the landmark date, gender, and race) were also included. 

\paragraph{Model Building and Evaluation.}
We constructed prediction models using data from a target year $t$ belonging to $\{2010,2011,\ldots,2020\}$ and, when applicable, source data from all preceding years ($\leq t-1$). 
Five modeling strategies were compared: (1) \textbf{ADAPT}; (2) a \textbf{Target-only} GLM trained on the current target data; (3) a GLM trained on \textbf{Pooled} data from all years up to $t$; (4) \textbf{TransGLM}; and (5) \textbf{Maximin}. We evaluated model performance under two settings: (i) current-to-current, where models are trained and evaluated on the same target year, and (ii) current-to-future, where models trained on the current year are evaluated on a future year. For each target year, data were partitioned into training and validation folds to perform cross-validation and assess transferability, with earlier years serving as source domains. Each target-year dataset was randomly split 40 times, and performance was summarized as the mean AUC across repetitions. 

To address class imbalance and reduce computational load, we used a case-control design that included all suicide attempt cases and randomly sampled 20 controls per case. 
Temporal generalizability was assessed by applying models trained on year $t$ to data from subsequent years, with predictive performance quantified by the AUC. We quantified model degradation over time by defining the \emph{aging effect} as the relative loss in AUC for a model trained $\Delta$ years earlier compared with a contemporaneous model trained and validated on the same target year $t$, averaged across available years:
\begin{equation}
    \mathrm{A}(\Delta)
= \frac{1}{T-\Delta} \sum_{t=\Delta+1}^T 
\frac{\mathrm{AUC}_t - \mathrm{AUC}_{t-\Delta}}{\mathrm{AUC}_t - 0.5}.
\label{eq: age effect}
\end{equation}

To assess cross-institutional generalization, models trained on MGB for different target years were transferred to Duke, where we conducted analogous ``current $\to$ current'' and ``current $\to$ future'' evaluations. We transport model trained on MGB year $t$ data to Duke year $t$ data for current $\to$ current evaluation; and to Duke year $t+\Delta$ data for current $\to$ future evaluation. We further quantified cross-institutional model degradation using $\mathrm{A}_{\text{cross-ins}}(\Delta)$ defined similar to \eqref{eq: age effect}, except that $\mathrm{AUC}_t$ denotes the AUC of the MGB-trained model evaluated on Duke data from year $t$, and $\mathrm{AUC}_{t-\Delta}$ corresponds to the model trained on MGB in year $t-\Delta$ but evaluated on Duke data from year $t$. Thus, $\mathrm{A}_{\text{cross-ins}}(\Delta)$ captures the relative performance loss of a \emph{transferred} model trained $\Delta$ years earlier compared with a contemporaneous transferred model evaluated on the same Duke target year.
To examine model stability over time, we summarized the distribution of coefficient estimates across different target years.

\section{Results}
\subsection{Results from Synthetic Data Study}
\begin{figure}[H]
    \centering
    \includegraphics[width=.9\linewidth]{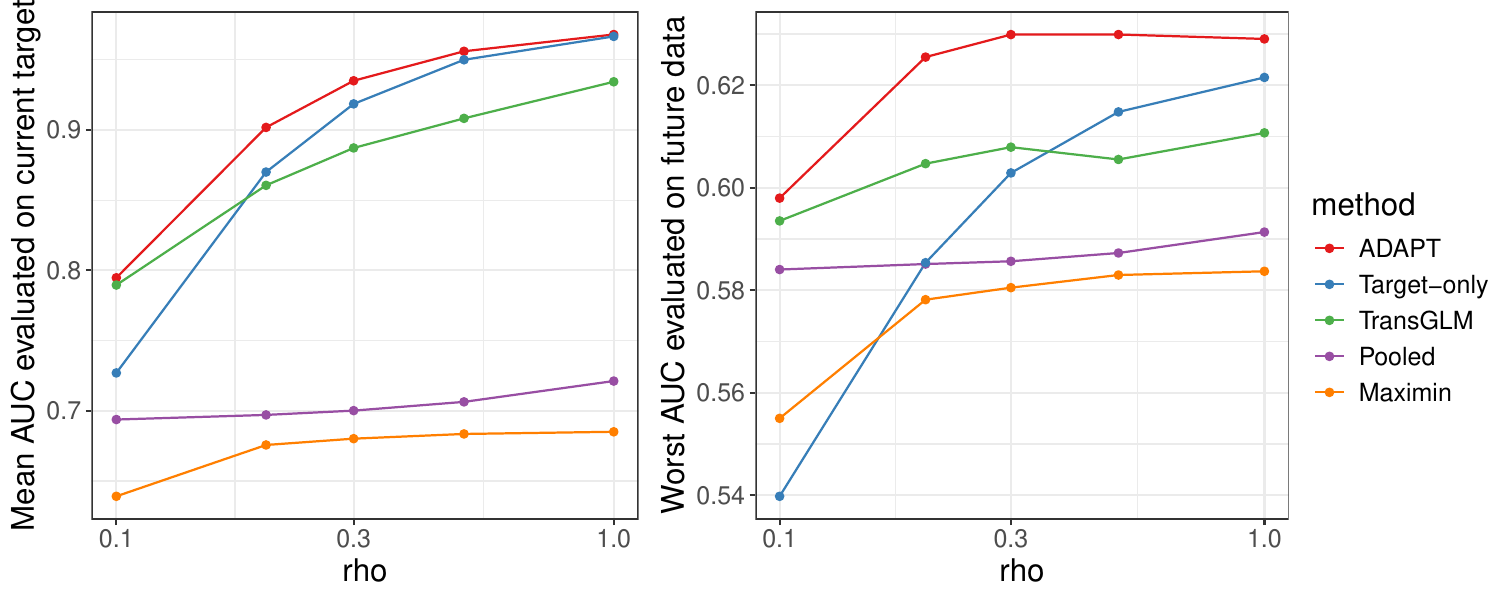}
    \caption{AUC as the current-to-historical data ratio increases. Left: validation data come from the same period as the the training data; right: validation data come from periods after the training data.}
    \label{fig:sim 1.1}
\end{figure}

\paragraph{Performance on the current target population.}
When validation data come from the same period as the training target (Figure \ref{fig:sim 1.1}, left), ADAPT consistently benefits from historical sources. Even with limited target data, i.e., ratio of the current data sample size to the historical sample size $\rho=0.1$, the performance of ADAPT matches transGLM and exceeds other baselines. As $\rho$ increases, ADAPT converges to the target-only estimator, while pooling and maximin remain insensitive to $\rho$ and consistently underperform. TransGLM performs comparably to ADAPT when $\rho=0.1$, but falls below target-only once $\rho \geq 0.3$.

\paragraph{Generalization to future populations.}
To assess robustness under temporal shift, we applied models trained at period $t$ to data from subsequent periods $t'>t$ and summarized performance by the worst AUC achieved across future periods (Figure \ref{fig:sim 1.1}, right). This captures how well models generalize once deployed on unseen data subject to distributional changes. ADAPT consistently attains the highest worst-case AUC on future data sets for all $\rho$, indicating greater stability under temporal shift. Maximin only surpasses target-only when $\rho=0.1$ but otherwise yields the lowest generalizability, while transGLM ranks second for $\rho \leq 0.3$ before falling below target-only. Pooling remains relatively stable but consistently below ADAPT and transGLM.

\begin{figure}[H]
    \centering
    \includegraphics[width=.75\linewidth]{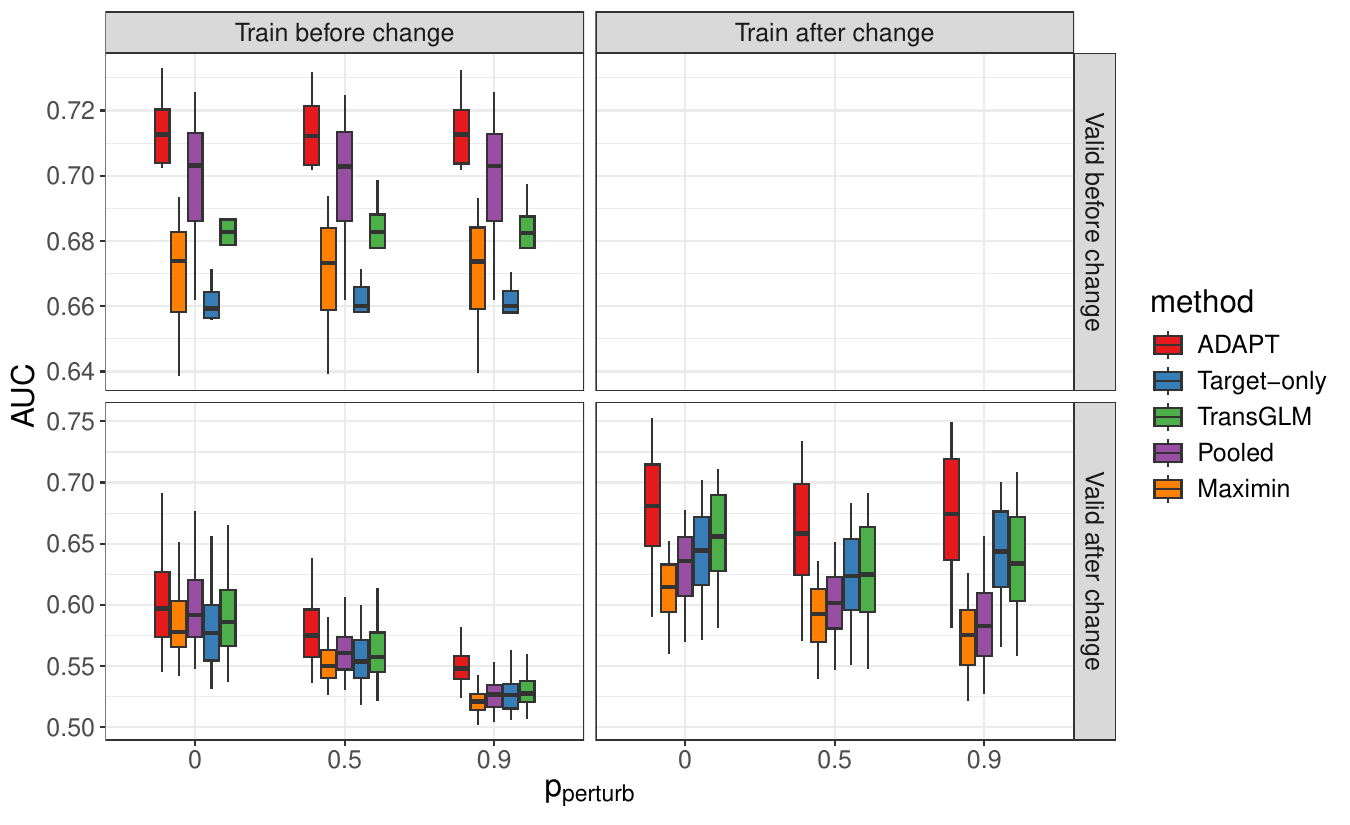}
    \caption{AUC as perturbation level increases. Time window for the training target and validation data is split into two groups: before perturbation ($t\leq 7$) and after perturbtion ($t\geq 8$)}
    \label{fig:sim 2.1}
\end{figure}

\paragraph{Generalization under perturbed design.}
We further assessed robustness by partitioning training and validation into pre-shift ($t \leq 7$) and post-shift ($t \geq 8$) periods (Figure \ref{fig:sim 2.1}). Evaluations within the same period type (pre–pre or post–post) yield higher AUC than those spanning the shift. Training on pre-shift data and validating on post-shift data shows the greatest degradation, which worsens with increasing $p_{\mathrm{perturb}}$, reflecting the difficulty of anticipating future changes. ADAPT deteriorates more slowly in this setting and maintains the highest medians, indicating robustness to post-shift distributions. Within-regime evaluations are less sensitive to $p_{\mathrm{perturb}}$, though pooling and maximin continue to decline even when trained and validated post-shift, suggesting limited forward generalizability. Pooling and transGLM perform competitively when both training and validation are pre-shift, but drop sharply when applied across pre- to post-shift periods. The target-only model shows consistently poor generalizability to post-shift data, while maximin remains uniformly low across all settings, reflecting over-conservativeness.

\subsection{Real-World Assessment of Suicide Risk Model Degradation}

\begin{figure}[H]
    \centering
    \includegraphics[width=1\linewidth,page=1]{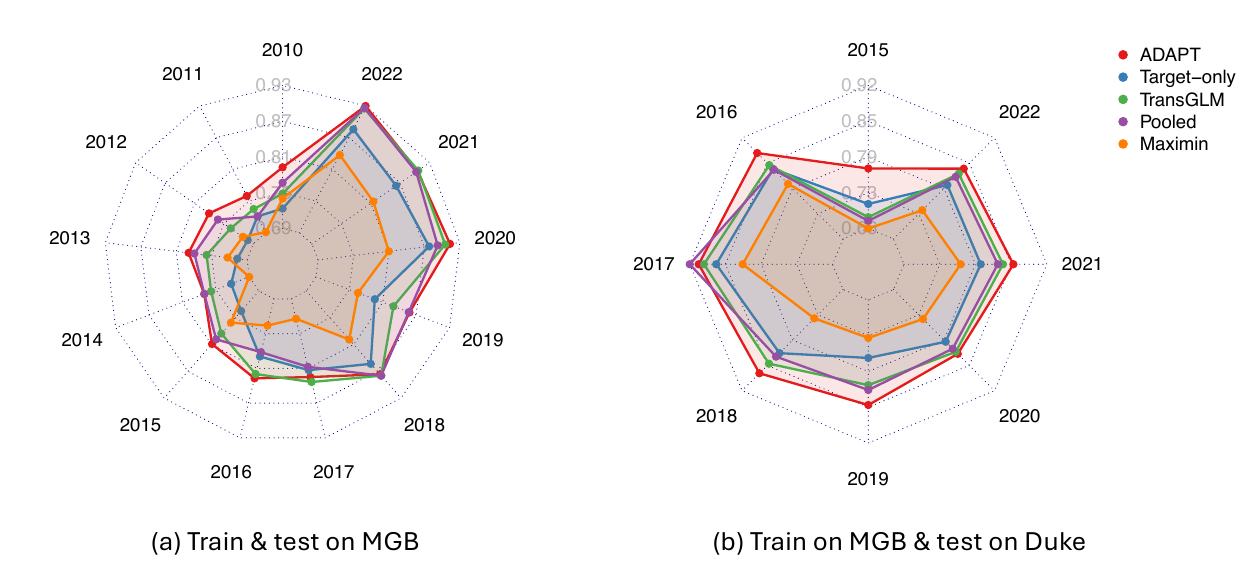}
    \caption{AUC when the training year and the validation year are the same.}
    \label{fig: real data trans}
\end{figure}

\paragraph{Accuracy on Current Target Data.} 
As shown in the left panel of Figure \ref{fig: real data trans}, when the training target in MGB data exactly matched the validation population (models trained on year $t$ and evaluated on the testing data from the same year $t$), ADAPT consistently achieved the best performance (mean AUC: 0.815), with especially pronounced gains in 2010–2015. Pooled logistic regression generally ranked second (mean AUC: 0.800), followed by transGLM with slightly worse average performance (mean AUC:0.794). The period-specific logistic model lagged, reflecting efficiency loss from ignoring historical data (mean AUC: 0.763). Maximin performed worst (mean AUC: 0.730), which suffers from overly conservative estimation in this setting, attenuating informative signal. Collectively, these results demonstrates ADAPT’s advantage in leveraging heterogeneous historical sources and transfer the knowledge to learn well for the current-target period, particularly when targeting earlier years. 

The right panel of Figure \ref{fig: real data trans} illustrates cross-site model generalizability based on same-year validation using Duke’s data. Consistent with the results on the MGB cohort, ADAPT demonstrated a clear performance gain compared to other methods when validated on the same-year but cross-site data (ADAPT: mean AUC 0.846). TransGLM and Pooled achieved the comparable cross-site transferrability which ranked the second and the third (TransGLM: mean AUC 0.821; Pooled: mean AUC 0.818). Target-only (mean AUC:  0.799), and maximin (mean AUC: 0.749) ranked fourth and fifth, respectively, when validated on Duke. All methods exhibited a modest performance increase when transferred from MGB to Duke, possibly due to different levels of data quality on MGB and Duke.

\begin{figure}[H]
    \centering
    \includegraphics[width=.9\linewidth,page=2]{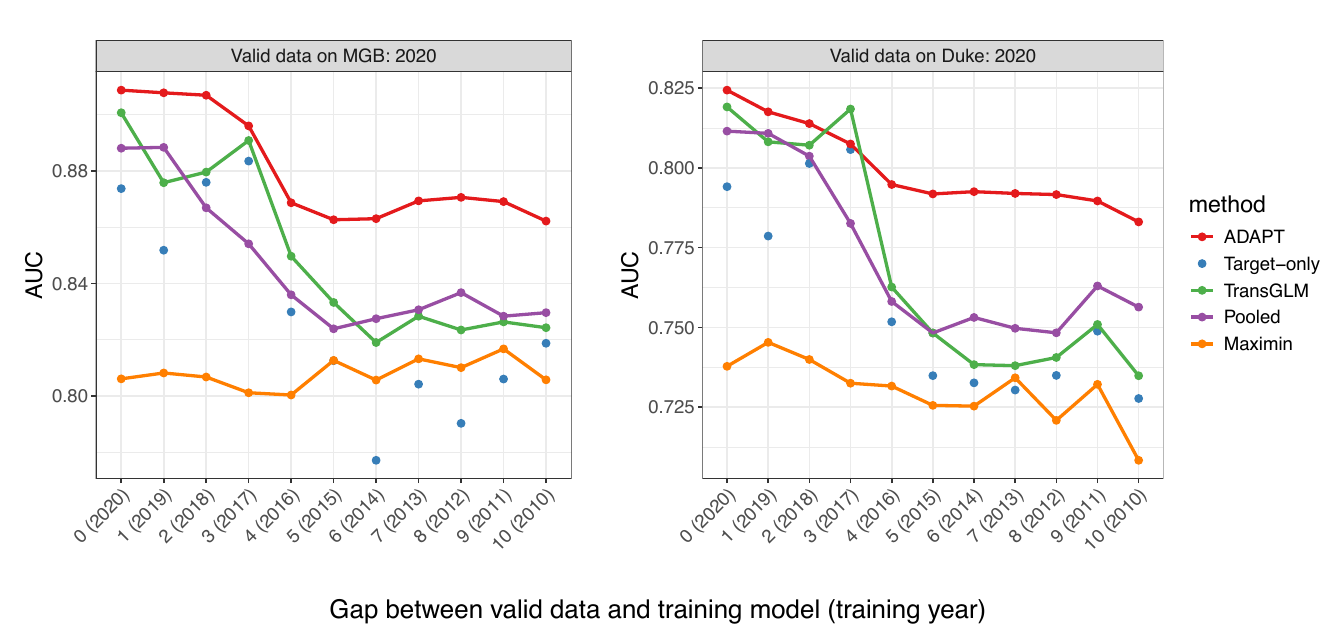}
    \caption{AUC when the validation data comes from 2020 while models are trained for a target period (shown on x-axis) prior to 2020.}
    \label{fig: real data gen}
\end{figure}

\paragraph{Long-Term Performance Maintenance.}
On the MGB site, we fixed 2020 as the validation year and trained on progressively earlier periods (left panel of Figure \ref{fig: real data gen}). As the train-test gap widens, AUC declines for all methods, but at different rates. ADAPT achieved the best long-term stability (mean AUC: 0.881, standard deviation (sd) over different training periods: 0.019), maintaining performance between 0.862 and 0.909 even with decade-old training data, and showing only a modest dip around the 2016 systemic shift (2016: 0.869 to 2015: 0.863). In contrast, transGLM (mean AUC: 0.850, sd: 0.031) exhibited larger drops (0.850 to 0.833). Simply pooling all historical data (mean AUC: 0.846, sd: 0.024) led to a clear degradation trend especially from pooling-till-2019 to pooling-till-2015 that drops AUC from 0.888 to 0.824 when validated on 2020's data. The target-only method showed the weakest temporal generalizability (mean AUC: 0.829, sd: 0.037), while maximin produced relatively flat but consistently low performance (mean AUC: 0.808, sd: 0.005), reflecting a overly conservative worst-case weighting strategy that misaligned with the actual direction of temporal drift. 

On the right panel of Figure \ref{fig: real data gen}, we transferred models trained on MGB targeting different time periods to Duke’s 2020 data to evaluate cross-site generalizability. ADAPT consistently achieved the highest AUC (mean AUC: 0.800, sd: 0.013), displaying a similar degradation pattern as the training year diverged further from the 2020 test year. Pooled (mean AUC: 0.771, sd: 0.026), TransGLM (mean AUC: 0.770, sd: 0.036) and Target-only (mean AUC: 0.758, sd: 0.031) exhibited a similar downward trend, whereas maximin maintained relatively stable performance (mean AUC: 0.730, sd: 0.010), consistent with its robustness observed on the MGB cohort.

\begin{figure}[H]
    \centering
    \includegraphics[width=1.1\linewidth,page=4]{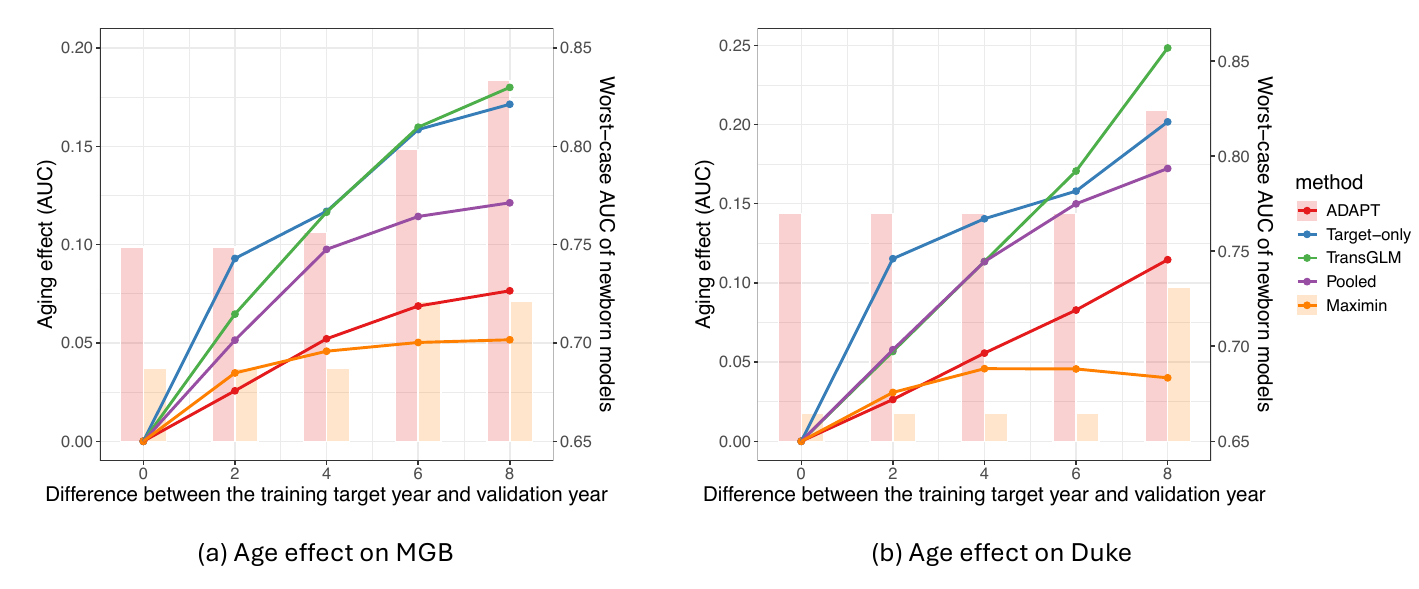}
    \caption{Left axis: age effect as a function of the year gap $\delta t$ between training and validation;
right axis: worst-case AUC of the newborn ADAPT and maximin model over different validation years within each $\delta t$ window.}
    \label{fig: real data age}
\end{figure}

To summarize the long-term performance across years, we define model aging as the relative AUC degradation of models trained $\Delta t$ years earlier compared with a ``newborn'' model trained in the same year (supposing that yearly retraining was possible). In the left panel of Figure \ref{fig: real data age}, where the age effect is estimated using the MGB cohort, ADAPT exhibits the smallest short-term performance loss (2-year average degradation: 2.57\%; 8-year average degradation: 7.65\%), while maximin shows the smallest long-term loss (2-year average degradation: 3.47\%; 8-year average degradation: 5.16\%), highlighting the relative stability of both methods over time. However, Maximin’s modest degradation is mainly attributable to its lower baseline AUC (barplot in Figure \ref{fig: real data age}), which consistently exhibits the lowest worst-case newborn-model AUCs across the $\delta t$ windows (8-year worst-case AUC: 0.721 for Maximin vs 0.833 for ADAPT), indicating weaker baseline performance rather than improved temporal robustness. Pooled shows an intermediate decline in both the short term (2-year average degradation: 5.14\%) and the long term (8-year average degradation: 12.1\%). In contrast, Target-only and TransGLM deteriorate the most, with degradation exceeding 17\% when the training models are 8 years older than the validation data.

When we externally evaluate the methods on the Duke cohort (right panel of Figure \ref{fig: real data age}), the age effects quantify performance relative to the "newborn" model trained on MGB and validated on Duke data from the same calendar year. Under this setting, ADAPT again exhibits the smallest short-term degradation (2-year degradation: 2.64\%), whereas Maximin shows the slowest long-term temporal deterioration (8-year degradation: 1.77\%) yet the poorest worst-case baseline performance (8-year worst-case AUC: 0.731 for Maximin). Relative to the MGB evaluation, TransGLM shows a comparable 2-year decline (5.67\% on Duke vs. 6.46\% on MGB) but substantially worse 8-year degradation (24.8\% on Duke vs. 18.0\% on MGB). Consistent with its rapid decay on MGB, the Target-year-only model has the largest short-term degradation on Duke (11.5\% at 2 years). Overall, ADAPT slows temporal degradation while preserving the current-period performance, extending model longevity under distributional shift and possible site heterogeneity.

\begin{figure}[H]
    \centering    \includegraphics[width=1\linewidth,page=1]{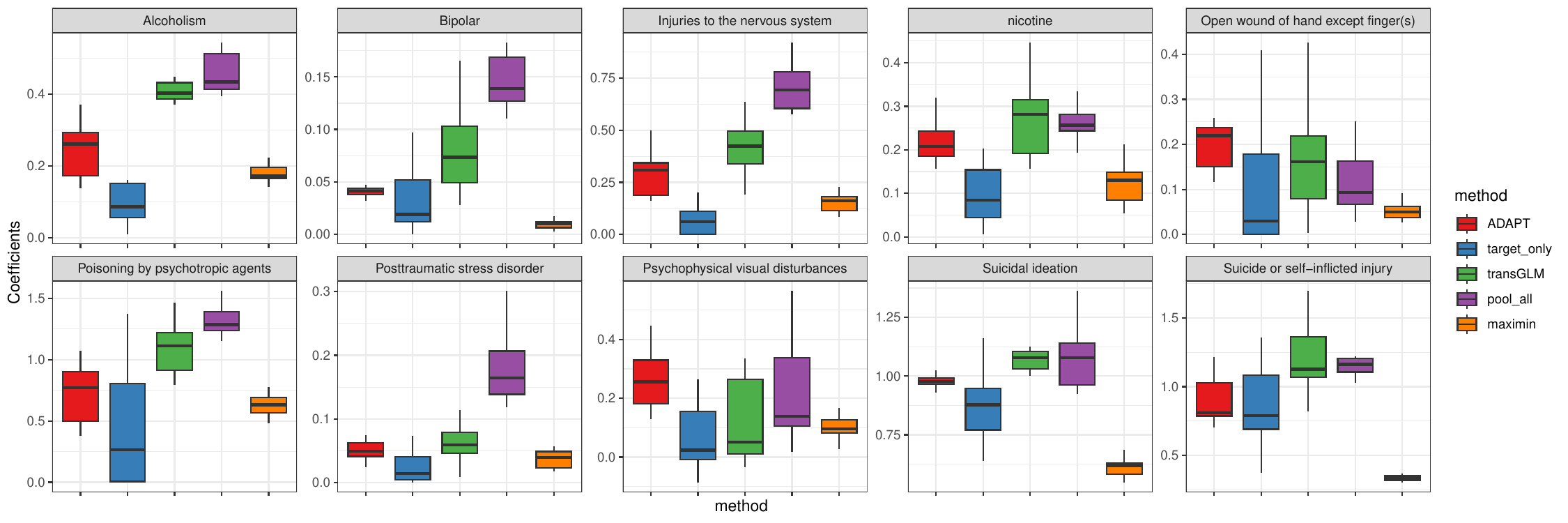}
    \caption{Boxplot summarizes the variation of the ten most influential features defined by the average absolute coefficient magnitude across years (narrower means less variant).}
    \label{fig: coef adapt}
\end{figure}

\paragraph{Model Interpretations.}
To understand ADAPT’s superior generalizability, we examined in Figure \ref{fig: coef adapt} the temporal stability of the effects of the ten most influential features, including \emph{alcoholism}, \emph{bipolar}, \emph{nervous system injury}, \emph{nicotine}, \emph{open hand wound}, \emph{poisoning by psychotropic agents}, \emph{PTSD}, \emph{psychophysical visual disturbances}, \emph{suicidal ideation}, and \emph{self-inflicted injury}. Features are identified by aggregating method-specific rankings of mean absolute coefficient magnitudes across years and retaining the ten highest-ranked unique ones. Narrower boxes indicate lower across-year standard deviation (sd) and higher temporal stability. ADAPT produced consistently tight inter-year distributions with non-zero magnitudes (average sd of coefficients over years: 0.086), suggesting it captures persistent signals. The pooled model showed the next highest stability (average sd: 0.106), likely due to its larger effective sample size. In contrast, target-only (average sd: 0.179) and transGLM (average sd: 0.114) displayed much wider dispersion, reflecting higher susceptibility to systemic shifts. Maximin yielded the narrowest boxes overall (average sd: 0.044) but at uniformly attenuated magnitudes, consistent with its over-conservative weighting that suppresses predictive signals. These patterns mirror the accuracy results: ADAPT combines stability with predictive strength, yielding the most reliable generalization to future data.

\section{Discussion}

We introduce ADAPT, a novel framework designed to enhance the longevity of clinical AI by proactively addressing temporal data shifts. By leveraging a robustness-aware learning approach, ADAPT optimizes the prediction performance under the most challenging future scenarios derived from historical and current data. This enables adaptive robustness: high accuracy for present patient groups while protecting against unforeseen changes over time. Such capability is especially vital in healthcare, where AI models are susceptible to performance decay due to temporal shifts, and failures can have serious real-world consequences. ADAPT’s effectiveness in suicide risk prediction illustrates its potential to improve the stewardship of clinical AI systems. The framework can be applied to other areas facing similar temporal shifts, from chronic disease management to epidemiological forecasting. Moreover, our learning pipeline could be adapted to support more complex neural network architectures. By transforming AI from a static, one-time tool into a continuously reliable asset, ADAPT addresses two major causes of real-world failure: overfitting to initial training conditions and underestimating future shifts. This increased reliability may strengthen trust in AI and accelerate its practical integration into healthcare. While ADAPT operates under the assumption that future shifts will generally follow patterns observed in the past, truly unprecedented events such as a new pandemic would require additional complementary strategies. Future work will focus on extending ADAPT to more complex machine learning architectures, integrating it with semi-supervised learning using large unlabeled datasets, and validating its generalizability across diverse healthcare systems. Through its proactive emphasis on accuracy, robustness, and longevity, ADAPT contributes to the development of sustainable and dependable clinical AI.

\addtolength{\textheight}{-0.7in}%
\addtolength{\topmargin}{0.4in}%

\bibliographystyle{apalike}
\bibliography{Bibliography-MM-MC}

\newpage
\appendix

\section*{Supplementary Material}

\section{Supplementary Details for ADAPT Algorithm}
\label{sec: supp a}

\subsection{Setup and Notations}
Let $Y$ denote a possibly binary, continuous, or count outcome variable, and $X$ represent the predictors that can be either observed risk factors or embeddings derived from some pretrained deep neural network for highly complex data. ADAPT assumes access to labeled data from $L$ source populations, denoted by $\Dsc\supl = (\YY\supl, \XX\supl) \in \mathbb{R}^{N_l \times p}$, and a target dataset $\Dsc\supQ = (\YY\supQ, \XX\supQ) \in \mathbb{R}^{n \times p}$. Each source may reflect data collected at a distinct historical time point and is modeled by a generalized linear model (GLM), $\Pbb\subYX(\bbeta\supl)$ with conditional mean $\Ebb(\YY\supl \mid \XX\supl) = g^{-1}\{(\XX\supl)\trans \bbeta\supl\}$, where $g(\cdot)$ is a known link function. The current target population is assumed to follow a separate GLM: $\Pbb\subYX(\bbeta\supQ)$ with parameter $\bbeta\supQ$, such that $\Ebb(\YY\supQ \mid \XX\supQ) = g^{-1}\{(\XX\supQ)\trans \bbeta\supQ\}$. Let $\Pbb\supQ\subX$ denote the covariate distribution of the current target population, which we assume remains unchanged in the future (i.e., no covariate shift). The conditional distribution of the future target $\Pbb\subYX(\bbeta\supTbb)$ is also assumed to follow the same GLM family, but with a possibly shifted parameter $\bbeta\supTbb \ne \bbeta\supQbb$. 

\subsection{ADAPT Algorithm}\label{sec:supp:a2}
The ADAPT method proceeds in three key steps:  
(i) estimation of source-specific and current target models;  
(ii) construction of an uncertainty set $\Csc(\tau)$ characterizing plausible future targets;  
(iii) optimization of a worst-case loss over the uncertainty set via a first-order approximation.  
The details of the steps will be introduced below. Importantly, each step requires only summary-level statistics from source datasets and does not rely on access to individual-level information. To avoid over-fitting and facilitate reliable estimation, the target data $\Dsc\supQ$ is randomly split into two disjoint subsets $\Dsc\supQ_1$ and $\Dsc\supQ_2$. The first half of the target data is used for model estimation, while the second half is used to evaluate performance when constructing a future model. This strategy is critical to avoid assigning disproportionate weights to the current-target-only estimator, which could occur if both estimation and evaluation relied on the same data. 
\paragraph{{Step 1:  Capturing Historical and Current Signals.}} We begin by estimating predictive models for the current target and each of the $L$ source populations. For each source $l$, we estimate its model parameter $\bbetahat\supP$ by minimizing the penalized negative log-likelihood 
$$\ell(\bbeta; \Dsc\supP) +  \Psc_{\lambda}(\bbeta),$$ 
on the source dataset $\Dsc\supP$, where $\ell(\cdot,\cdot)$ is the negative log-likelihood function and $\Psc_{\lambda}(\cdot)$ is a penalty function such as the elastic net specified by the user, with $\lambda$ being a tuning parameter that can be selected via standard procedures including cross-validation. To estimate the target model, we randomly split the target dataset into two halves. Using the first half, $\Dsc\supQ_1$, we obtain $\bbetahat\supQbb_1$ by the same penalized likelihood approach. 
This collection of estimated models forms the basis for constructing candidate future models. Note that although we consider the GLM here, our framework accommodates more sophisticated neural network methods for which $X$ can be naturally taken as the last layer of a pretrained neural network.

\paragraph{{Step 2: Defining the Horizon of Plausible Futures.}} We then construct an uncertainty set of plausible future models based on convex combinations of the estimated source and target models. Let $\BBhat = [\bbetahat\supQbb_1, \bbetahat^{(1)}, \dots, \bbetahat\supL]$ and consider candidate parameters of the form $\BBhat \gamma$ with $\gamma \in \Delta_{L+1}$. 
To ensure that these candidates remain compatible with the observed target, we require that they achieve adequate fit on the held-out portion of target data $\Dsc\supQbb_2$. 
Specifically, the derived uncertainty set is defined as
\[
\Csc(\tau) := \left\{ \bbeta\supTbb = \BBhat \gamma : \gamma \in \Delta_{L+1}, \; \ell(\BBhat \gamma; \Dsc\supQbb_2) \leq \tau \right\}.
\]

Here, $\tau \geq 0$ serves as a tuning parameter that balances the focus on the current target against robustness to future shifts. It controls the breadth of $\Csc(\tau)$: a small $\tau$ restricts $\Csc(\tau)$ to candidates that closely match the current target distribution, whereas a larger $\tau$ allow the model to account for broader deviations from the current target. 
To make a reasonable choice on $\tau$, we consider two potential issues in our data: (a) the target sample size is small so the current target estimator $\bbetahat\supQ_1$ itself is noisy; and (b) the historical sources exhibit substantial deviation from the current target, indicating that stronger potential drift can also happen for the future. It is important to note that under both (a) and (b), one should enlarge $\Csc(\tau)$ to guard against overly optimistic reliance on the current target model and improve out-of-sample robustness to plausible drift scenarios. Based on this intuition, we set
\[
\tau =\frac{1}{2}\big[\ell(\bbetahat\supQ_1; \Dsc_2\supQ)+\ell(\bbetatilde; \Dsc_2\supQ)\big],
\]
where $\bbetatilde$ is the convex combination of the source model coefficients optimizing the likelihood of the current target data. Again, to avoid over-fitting, we use the held-out sample $\Dsc_2\supQ)$ to construct likelihood loss $\ell$ in $\tau$. 

Since $\ell(\bbetahat\supQ_1; \Dsc_2\supQ)$ can be viewed as a performance measure of the target-only $\bbetahat\supQbb_1$, a large $\ell(\bbetahat\supQ_1; \Dsc_2\supQ)$ indicates the existence of issue (a). Similarly, a large $\ell(\bbetatilde; \Dsc_2\supQ)$ is a signal of issue (b). Thus, our choice of $\tau$ can serve for our purpose by automatically enlarging the uncertainty set $\Csc(\tau)$ to regularize the estimator when (a) or (b) occurs. We take an average of on the two likelihood loss function in $\tau$ to match its scale with $\ell(\BBhat \gamma; \Dsc\supQbb_2)$ in $\Csc(\tau)$. Also, this formulation implicitly imposes that any plausible future target model should perform no worse on the held-out target data than either the current target estimator or the best source-model combination. With such a proper choice of $\tau$, our construction avoids overly conservative extrapolation while ensuring that $\Csc(\tau)$ remains non-empty and informative.

\paragraph{{Step 3: Adversarial Optimization for Robustness.}}
Finally, ADAPT trains a model by minimizing the worst-case loss across the uncertainty set $\Csc(\tau)$. The objective is
\begin{equation}
\min_{\bbeta} \; \max_{\bbeta\supTbb \in \Csc(\tau)} \; 
\Ebb_{\Dsc\supTbb \sim \PP(\bbeta\supTbb)}
\big[ \ell(\bbeta; \Dsc\supTbb) - \ell(\bbeta\supinit; \Dsc\supTbb) \big],
\label{eq:adapt_def}
\end{equation}
where $\Dsc\supTbb$ denotes future target data drawn from $\PP(\bbeta\supTbb) := \{(\PP\subX\supQbb, \PP\subYX(\bbeta\supTbb))\}$ and $\bbeta\supinit$ is an initial estimator serving as a baseline to be discussed later. By constructing $\PP(\bbeta\supTbb)$ in this way, we actually assume the future temporal shift occurs in the conditional models parameterized by $\bbeta\supTbb$, while the covariate distribution remains fixed at the current target distribution $\PP\subX\supQbb$. The design of the worst-case objective in (\ref{eq:adapt_def}) helps seeking a robust model that performs no worse than the initial estimator $\bbeta\supinit$ under the most adversarial yet plausible outcome distribution within $\Csc(\tau)$.

In general, introducing an initial estimator $\bbeta\supinit$ provides a prior reference for the future target, and the objective explicitly focuses on improving robustness beyond this baseline for the worst-case future scenarios in $\Csc(\tau)$. In our studies, we choose $\bbeta\supinit$ to be the current-target-only estimator $\bbetahat\supQbb_1$. This choice plays two complementary roles. First, it provides a reasonable guess for the near-future target, and the excess loss in \eqref{eq:adapt_def} guarantees improved generalizability relative to the current target. In this sense, using $\bbetahat\supQbb_1$ as the baseline serves as an implicit regularization, discouraging solutions that sacrifice present-target performance without delivering gains under adverse future drifts in $\Csc(\tau)$. Second, using $\bbetahat\supQbb_1$ as the reference helps simplify the adversarial objective as will be introduced below. Alternatively, when one expects well-performed and generalizable source models, $\bbeta\supinit$ can be set as the convex combination of sources that best approximates the current target.

Directly optimizing the ADAPT estimator is computationally challenging due to the non-convexity of the adversarial objective function with respect to $\bbeta\supTbb$. To address this, we approximate $\ell(\bbeta, \bbeta\supTbb)$ via a second-order Taylor expansion around the initial model $\bbeta\supinit$:
\[
\Ebb_{\Dsc\supTbb \sim \PP(\bbeta\supTbb)}
\big[ \ell(\bbeta; \Dsc\supTbb) - \ell(\bbeta\supinit; \Dsc\supTbb) \big]\approx (\bbeta\supinit - \bbeta)\trans 
\HH(\bbeta\supinit) 
(\bbeta\supinit - \bbeta),
\]
where $\HH(\bbeta\supinit)$ is the Hessian of the log-likelihood function evaluated at $\bbeta\supinit$. This yields a tractable quadratic approximation while preserving robustness guarantees, which can be further equivalently converted to a single minimization problem with an explicit-form solution:
\begin{equation}
\bbeta_{\text{ADAPT}} = 
\argmin_{\bbeta\in \Csc(\tau)} \; 
(\bbeta\supinit - \bbeta)\trans 
\HH(\bbeta\supinit) 
(\bbeta\supinit - \bbeta),
\label{eq:adapt_est}
\end{equation}
which produces our final estimator $\bbeta_{\text{ADAPT}}$. The quadratic-form expression also reveals that the ADAPT estimator can also be interpreted as the projection of $\bbeta\supinit$ onto a restricted space that balances prediction accuracy on the current target and robustness to future distributional shifts. 

\section{Supplementary Details for Validation Studies}

\subsection{Simulation Design}\label{sec:supp:sim}
At discrete time points $l \in\{ 1, \dots, L = 15\}$, we observed independent datasets $\{\XX\supl, \YY\supl\}$. Each feature vector was generated as 
$X_{i}\supl \sim \mathcal{N}_p(\mathbf{0}, I_p)$ with $p = 100$, and outcomes were sampled from a logistic model with time-specific coefficients $\bbeta\supl$:
\[
P(Y\supl_i = 1) = g(\{X\supl_i\}\trans \bbeta\supl).
\]
The coefficient path $\{\bbeta\supl:l=1,\ldots,L\}$ evolved smoothly according to an autoregressive mixing process with occasional sparse perturbations to simulate abrupt drifts. These two mechanisms are designed to mimic realistic sources of distribution shift including gradual changes (e.g., evolving disease prevalence over time) and abrupt changes (e.g., coding-system updates). Initialization began with $m$ (the autoregressive order) sparse seed vectors drawn i.i.d.\ from $\mathcal{N}(0,1)$, with $p_0 = 30$ coordinates set to zero. At each subsequent time point, coefficients were updated via the autoregressive mechanism with stochastic shocks, except at $l=8$, where a global perturbation was introduced to emulate a major regime change: 
\[
\bbeta\supl =
\begin{cases}
\displaystyle 
\Big(\sum_{t=l-m}^{l-1} \gamma_t \bbeta^{(t)} \Big) \odot (1-\mathbf{v}_l) 
+ \mathcal{N}(0, \sigma_{\mathrm{shock}}^2) \odot \mathbf{v}_l, 
& l \neq 8, \\[1.5ex]
\displaystyle 
(1-p_{\mathrm{perturb}})
\Big[ \Big(\sum_{t=l-m}^{l-1} \gamma_t \bbeta^{(t)} \Big) \odot (1-\mathbf{v}_l) 
+ \mathcal{N}(0, \sigma_{\mathrm{shock}}^2) \odot \mathbf{v}_l \Big] 
+ p_{\mathrm{perturb}}\, \mathcal{N}(0, 0.5 I), 
& l = 8 .
\end{cases}
\]
where ``$\odot$'' represents the Hadamard (element-wise) product of two vectors. Here, each element $v_{jl}$ is drawn from a Bernoulli($p_{\mathrm{shock}}$) distribution with $p_{\mathrm{shock}} = 0.2$, indicating whether the $j$-th coefficient undergoes a sudden random change.

\subsection{Cohort characteristic}\label{sec: supp demo}

\begin{table}[!htbp]
\centering
\caption{Baseline characteristics of the Duke and MGB cohorts. Values are shown as $n$ (\%).}
\label{tab:cohort_characteristics}
\scriptsize
\setlength{\tabcolsep}{4pt}
\renewcommand{\arraystretch}{1.05}
\begin{tabular}{lccc}
\toprule
 & Duke (N=235,097) & MGB (N=412,943) & Overall (N=648,040) \\
\midrule
\multicolumn{4}{l}{\textbf{Gender}}\\
Female & 155252 (66.0\%) & 266318 (64.5\%) & 421570 (65.1\%) \\
Male & 79845 (34.0\%) & 146625 (35.5\%) & 226470 (34.9\%) \\
\addlinespace
\multicolumn{4}{l}{\textbf{Race}}\\
Asian & 4942 (2.1\%) & 11008 (2.7\%) & 15950 (2.5\%) \\
Black & 50538 (21.5\%) & 24682 (6.0\%) & 75220 (11.6\%) \\
White & 165785 (70.5\%) & 328755 (79.6\%) & 494540 (76.3\%) \\
Other & 13832 (5.9\%) & 47367 (11.5\%) & 61199 (9.4\%) \\
Unknown & 0 (0\%) & 1131 (0.3\%) & 1131 (0.2\%) \\
\addlinespace
\multicolumn{4}{l}{\textbf{Ethnicity}}\\
Hispanic & 10771 (4.6\%) & 17330 (4.2\%) & 28101 (4.3\%) \\
Non-Hispanic & 216330 (92.0\%) & 395613 (95.8\%) & 611943 (94.4\%) \\
Unknown & 7996 (3.4\%) & 0 (0\%) & 7996 (1.2\%) \\
\addlinespace
\multicolumn{4}{l}{\textbf{Age group}}\\
0-18 & 20734 (8.8\%) & 27872 (6.7\%) & 48606 (7.5\%) \\
19-25 & 18879 (8.0\%) & 35919 (8.7\%) & 54798 (8.5\%) \\
26-55 & 104991 (44.7\%) & 193528 (46.9\%) & 298519 (46.1\%) \\
\textgreater{}55 & 90493 (38.5\%) & 155624 (37.7\%) & 246117 (38.0\%) \\
\addlinespace
\multicolumn{4}{l}{\textbf{Landmark year}}\\
2005 & 0 (0\%) & 10029 (2.4\%) & 10029 (1.5\%) \\
2006 & 0 (0\%) & 10859 (2.6\%) & 10859 (1.7\%) \\
2007 & 0 (0\%) & 11259 (2.7\%) & 11259 (1.7\%) \\
2008 & 0 (0\%) & 11980 (2.9\%) & 11980 (1.8\%) \\
2009 & 0 (0\%) & 13355 (3.2\%) & 13355 (2.1\%) \\
2010 & 0 (0\%) & 14380 (3.5\%) & 14380 (2.2\%) \\
2011 & 0 (0\%) & 15355 (3.7\%) & 15355 (2.4\%) \\
2012 & 0 (0\%) & 17128 (4.1\%) & 17128 (2.6\%) \\
2013 & 0 (0\%) & 17854 (4.3\%) & 17854 (2.8\%) \\
2014 & 11498 (4.9\%) & 20125 (4.9\%) & 31623 (4.9\%) \\
2015 & 16245 (6.9\%) & 21335 (5.2\%) & 37580 (5.8\%) \\
2016 & 18370 (7.8\%) & 25714 (6.2\%) & 44084 (6.8\%) \\
2017 & 20849 (8.9\%) & 33165 (8.0\%) & 54014 (8.3\%) \\
2018 & 21734 (9.2\%) & 36727 (8.9\%) & 58461 (9.0\%) \\
2019 & 23673 (10.1\%) & 40224 (9.7\%) & 63897 (9.9\%) \\
2020 & 23656 (10.1\%) & 38474 (9.3\%) & 62130 (9.6\%) \\
2021 & 26194 (11.1\%) & 42218 (10.2\%) & 68412 (10.6\%) \\
2022 & 26689 (11.4\%) & 32762 (7.9\%) & 59451 (9.2\%) \\
2023 & 26574 (11.3\%) & 0 (0\%) & 26574 (4.1\%) \\
2024 & 19040 (8.1\%) & 0 (0\%) & 19040 (2.9\%) \\
2025 & 575 (0.2\%) & 0 (0\%) & 575 (0.1\%) \\
\bottomrule
\end{tabular}
\end{table}

\end{document}